\documentclass[a4paper,fleqn,usenatbib]{aa}

\usepackage{graphicx}
\usepackage{txfonts}
\usepackage{times}
\usepackage{booktabs}
\usepackage{caption}
\usepackage{multirow}
\usepackage{multicol}
\usepackage{amsmath,amssymb}
\usepackage[breaklinks=true,linkcolor=blue,allcolors=blue,colorlinks=true]{hyperref}

\newcommand{\Mstar}{\ensuremath{\mathrm{log\,(M_\star/M_\odot)}}}
\begin{document}

   \title{Radio properties of Green Pea galaxies}

   \subtitle{}

   \author{A. Borkar
          \inst{1},
          R. Grossov\'{a}\inst{1,2},
          J. Svoboda\inst{1},
          E. Moravec\inst{3},
          K. Kouroumpatzakis\inst{1},
          P. G. Boorman\inst{4},\\
          B. Adamcov\'{a}\inst{1},
          B. Mingo\inst{5},
          M. Ehle\inst{6}
          }
    \authorrunning{Borkar et al.}
   \institute{$^1$ Astronomical Institute of the Czech Academy of Sciences, Bo\v{c}n\'i II 1401, CZ-14100 Prague, Czech Republic,
              \email{borkar@asu.cas.cz}.\\
              $^2$ Department of Theoretical Physics and Astrophysics, Faculty of Science, Masaryk University, Kotl\'{a}\v{r}sk\'{a} 2, Brno, CZ-61137, Czech Republic.\\
              $^3$ Green Bank Observatory, P.O. Box 2, Green Bank, WV 24944, USA\\
              $^4$ Cahill Center for Astronomy and Astrophysics, California Institute of Technology, Pasadena, CA 91125, USA\\
              $^5$ School of Physical Sciences, The Open University, Walton Hall, Milton Keynes, MK7 6AA, UK\\
              $^6$ European Space Agency, European Space Astronomy Centre (ESA/ESAC), Camino Bajo del Castillo s/n, 28692 Villanueva de la Ca\~nada, Madrid, Spain}

   \date{Received September 15, 1996; accepted March 16, 1997}


  \abstract
   {}
   {Green Peas (GPs) are young, compact, star-forming dwarf galaxies, and local (\(z\sim0.3\)) analogues of the early galaxies (\(z\geq 6\)) considered to be mainly responsible for the reionisation of the Universe. Recent X-ray observations of GPs have detected high excess emission which cannot be accounted for by star formation alone, and implies presence of an active galactic nucleus (AGN). We employ radio observations to study the radio properties of GPs, build their radio spectral energy distributions, and verify the presence of AGNs.}
   {We performed new radio observations of three GPs with the Karl G. Jansky Very Large Array (JVLA) in the L, C, and X bands (1.4, 6 and 10 GHz resp.), supplemented by the data from archival observations and large radio surveys. We also analysed the archival radio data for a larger sample of GPs and blueberry (BBs) galaxies, which are lower-mass and lower-redshift analogues of the GPs. To understand the significance of the radio observations, we assess the detectability of these sources, and compare the detected radio luminosities with the expectations from theoretical and empirical relations.}
   {Two of the three targeted GPs were strongly detected (\(>10\sigma\)) in the JVLA observations and their fluxes are consistent with star formation, while the third source was undetected. Although a large fraction (\(\sim 75\%\)) of the sources from the larger archival sample of GPs and BBs could be detected with archival surveys, only a small number (\(< 40\%\)) are detected and their radio luminosity is significantly lower than the expectation from empirical relations.}
   {Our results show that the majority of the dwarf galaxy sample is highly underluminous. Especially towards the lower end of galaxy mass and star formation rate (SFR), the radio luminosity \textendash~SFR relation deviates from the empirical relations, suggesting that the relations established for larger galaxies may not hold towards the low-mass end.}


   \maketitle
%

\section{Introduction}

In the hierarchical formation model, larger galaxies (\(\Mstar\geq 10\)) observed at low redshift are formed through a continuous merging of smaller galaxies (see e.g. \citealt{Cole00}). The first galaxies were young, low-mass (dwarf, \(\Mstar <9.5\)) sources and were likely the hosts of the first generation of stars which were responsible for the reionisation of the Universe, when the neutral gas was almost completely ionised. This Epoch of Reionisation (\(6 < z < 20\); \citealt{Bouwens15, Robertson15}) is pivotal to our understanding of cosmology and structure formation. The dwarf galaxies present at this epoch are characterised by their low mass, vigorous star formation and relatively low metallicity, and their production of a vast amount of ultraviolet (UV) and Lyman continuum radiation from young, massive stars, which can ionise hydrogen atoms in the intergalactic medium \citep{Steidel01}. The intense star formation in these early galaxies is long thought to be the main cause of the cosmic reionisation \citep{Shapiro87,Loeb01}. Quasars and active galactic nuclei (AGN) \textendash~powered by accretion onto supermassive black holes \textendash~could also provide the necessary photons for the reionisation. The proportion of the contribution of star formation and AGN emission to cosmic reionisation is still under debate, and some studies have suggested that the AGN emission could substantially contribute to the reionisation \citep{Volonteri09,Giallongo15}.

\begin{table*}[htb]
    \centering
    \caption{Properties of the sample of Green Pea galaxies.}
    \label{tab:GPs}
    \begin{tabular}{lccccccc}
        \toprule
        \toprule
        Source & SDSS & RA & DEC & \(z\) & \(\mathrm{M}_\star\) & SFR & log L\(_\mathrm{X}\)\\
         name & name & [h:m:s] & [d:m:s] & & [log M\(_\odot\)] & [M\(_\odot\,\mathrm{yr}^{-1}\)] & \(\mathrm{erg\,s^{-1}}\)\\[1mm]
         \midrule\\[-3mm]
         {GP1} & J074936.77+333716.3 &  {07:49:36.773} & {+33:37:16.395} & {0.2733} & {9.8} & {58.8} & 42.08 \\[1mm]
         {GP2} & J082247.66+224144.0 &{08:22:47.672} & {+22:41:44.101} & {0.2163} & {9.6} & {37.4} & 42.07 \\[1mm]
         {GP3} & J133928.30+151642.1 & {13:39:28.309} & {+15:16:42.022} & {0.1920} & {9.3} & {18.8} & $< 41.11$\\
         \bottomrule
    \end{tabular}
\end{table*}

Until recently, these early galaxies were difficult to observe due to their high redshift, as the UV-optical emission from star formation has been shifted to infrared (IR) band, which is unreachable with ground-based observatories and the limited sensitivity of space observatories. With the advent of \textit{JWST}, redshift beyond \(\mathrm{z > 7}\) have become accessible (see e.g. \citealt{Naidu22,Castellano22,Finkelstein22,Adams23,Atek23b,Atek23a,Labbe23}). Some of these galaxies seem to be hosting AGNs \citep{Juodzbalis23, Bogdan23, Goulding23}.
Although the \textit{JWST} will uncover many high-z dwarf galaxies, studying them in detail is challenging as it requires long exposures, which may not be achievable for multi-wavelength observations. Thus, to get a holistic understanding of the physical processes in the galaxies responsible for cosmic reionisation, it is crucial to study their low-z analogues with multi-wavelength observations, which can be achieved with the current generation of ground-based and space observatories.

Green Pea galaxies (GPs) are a class of low-redshift (\(z \sim 0.3\)) galaxies discovered in the Sloan Digital Sky Survey (SDSS) Galaxy Zoo project \citep{Cardamone09}. GPs are compact (\(\leq 5\) kpc), low stellar mass (\Mstar~$\sim$ 9), have a high star formation rate (\(\mathrm{SFR \sim 10\,M_\odot\,yr^{-1}}\)), consequently very high specific SFR (defined as the ratio of galaxy stellar mass and SFR: \(\mathrm{sSFR = M_\star/SFR \sim 10^{-8}\, yr^{-1}}\)), and sub-solar metallicities (\(\mathrm{12+log[O/H] < 8.7}\)). Strong Lyman-\(\alpha\) lines are detected in a large fraction of GPs showing similarities with the high redshift compact star forming galaxies called the Lyman Alpha Emitters \citep{Hayes15,Yang17GPs}. These similarities with the high-z galaxies make the GPs ideal analogs to study the nature of the ionising radiation in detail \citep{Scaerer22, Rhoads23}.

Following the discovery of GPs, a few local (\(z < 0.1\)) samples of GP-like dwarf galaxies have been presented in the literature. \citet{Yang17} presented a sample of GP analogs (\(z<0.05\)), called ``blueberries", with much lower stellar masses (\Mstar~\(\approx {6.5} - {7.5}\)) and lower metallicities (\(7.1 < 12 + \mathrm{log(O/H)} < 7.8\)) but a similar or higher sSFR than GPs. The sample was selected using SDSS imaging and photometry, along with compactness and large equivalent widths of optical emission lines. A similar sample was presented by \citet{McKinney19} and \citet{Jaskot19}, who focused their selection on galaxies with the highest ionisation (high \ion{O}{III}/\ion{O}{II} line ratio). We call both these samples as `blueberries' (BBs) for simplicity.

\citet{Svoboda19} reported on the first X-ray observations of three GPs, using the \textit{XMM-Newton} observations of three GPs. The GPs were selected from the \citet{Cardamone09} sample of 80 sources, which were classified as purely star forming galaxies in the BPT diagram \citep{Baldwin81}, which had high SFR and lower-redshift in order to facilitate favourable observability with \textit{XMM-Newton}. These three selected sources are referred to by \citet{Svoboda19} as GP1, GP2 and GP3\footnote{The physical properties and X-ray observations are discussed in \citet{Svoboda19}.}. The targeted sources are at redshift \(z = 0.19 - 0.27\), have high SFR \(\sim 20 - 60\,\mathrm{M_\odot\,yr^{-1}}\). Their stellar masses (\Mstar~\(= 9.3 - 9.8\)) and metallicities (\(\mathrm{12 + log(O/H)} = 8.1 - 8.3\)) are slightly above average compared to the parent sample. The details of the three sources are in Table~\ref{tab:GPs}.

The X-ray observations with \textit{XMM-Newton} detected extreme X-ray luminosities for star forming galaxies of \(L_{\mathrm{2-10\,keV}} \sim 10^{42}\,\mathrm{erg\,s^{-1}}\) in two of the three targets (GP1 and GP2), a factor \(> 5\) larger than the estimated luminosity from theoretical or empirical relations for star forming galaxies of similar mass range (see \citealt{Svoboda19} for the details). At the same time, the third target, GP3, was entirely undetected, even in the deeper follow-up observations. In the BPT diagram \citep{Baldwin81}, GP1 is classified as strictly star forming, while GP2 and GP3 show line ratios which are on the border between the AGN-like and star forming classification \citep{Cardamone09}. The observed excess X-ray emission in the two GPs could not be explained by a higher efficiency of high mass X-ray binaries in the low metallicity environment \citep{Svoboda19} or contributions from hot gas \citep{Franeck22}, and it implies the presence of an AGN.

A sizable fraction (\(> 10 - 30\%\)) of dwarf galaxies are expected to host AGN \citep{Miller15,Kaviraj19,Reines22}. Radio observations provide a unique avenue in the search for AGN in dwarf galaxies, as radio emission can probe black holes accreting at low Eddington ratio, while the optical tracers tend to select higher accretion sources in low SFR galaxies \citep{Koerding06,Best12,Moravec22}. Recently, several AGN candidate sources have been detected in dwarf galaxies using radio observations \citep{Nyland12,Reines14,Latimer19,Mezcua19,Reines20}. Radio observations also provide an independent way of estimating the SFR in these galaxies. Motivated by this, we proposed targeted observations of the three GPs with JVLA.

In this paper, we present the results of our dedicated JVLA observations of the three GPs, along with the analysis of the archival radio data for three GPs. We also explore the radio properties of the full GP sample from \citet{Cardamone09} and the BB sample from \citet{Yang17} and \citet{Jaskot19} as a comparison.
The paper is arranged in the following way:
In Sect.~\ref{sec:Data Analysis}, we present the details of the JVLA observations, data reduction and the observational results. Further, we present the compilation of the archival radio data for the targeted GPs and the samples of dwarf galaxies. We compare the radio observations of our targeted GPs with the comparison sample in Sect.~\ref{sec:Comparison}. In Sect.~\ref{sec:SFR-relation}, we explore the relation between star formation and radio flux in dwarf galaxies and its implications to the observations of GPs. The main conclusions are summarised in Sect.~\ref{sec:Conclusions}.
We use the following cosmological parameters: \(H_0 = 67.7\,\mathrm{km\,s^{-1}\,Mpc^{-1}}\), \(\Omega_M = 0.31\), \(\Omega_\Lambda = 0.69\) \citep{Planck20}. For radio spectral index \(\alpha\), we use the convention \(S_\nu \propto \nu^{\alpha}\).

\begin{table}
    \begin{center}
        \caption{VLA observation parameters {for the sources with detections}. The RMS noise in the images is \(\sim 7\,\mu\)Jy\,beam\(^{-1}\).}
        \label{tab:VLA_obs}

        \begin{tabular}{lcccc}
        \toprule
        \toprule
        Source & Time & Effective & Total & Beam\\
        & on source & Frequency & flux & size\\
        & [min] & [GHz] & [\(\mu\)Jy] & [\(''\times''\)]\\[1mm]
        \midrule\\[-3mm]
        \multirow{2}{*}{GP1} & 14.2 & 6.0 & 220\(\pm\)6 & 1.0 \(\times\) 0.9 \\
        & 14.5 & 10.0 & 142\(\pm\)10 & 0.6 \(\times\) 0.6 \\[3mm]
        \multirow{2}{*}{GP2} & 13.4 & 6.0 & 189\(\pm\)11 & 1.2 \(\times\) 0.9\\
        & 13.9 & 10.0 & 102\(\pm\)9 & 0.6 \(\times\) 0.5\\[1mm]
        \bottomrule
    \end{tabular}
    \end{center}
\end{table}

\section{Radio data analysis}\label{sec:Data Analysis}

\subsection{JVLA observations}

\begin{figure*}[htb]
    \centering
    \includegraphics[width=0.75\textwidth]{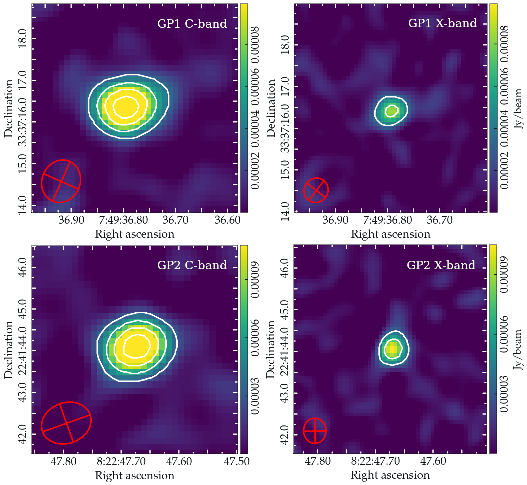}
    \caption{The completed JVLA observations of the GPs showing detections of GP1 and GP2 in the C- and X-bands. The red ellipses in the bottom-left corner indicate the beam size and shape. The white contours represent 1\(\sigma \times (5, 9, 13, 17)\) where 1\(\sigma = 7\,\mu\)Jy.}
    \label{fig:GP-obs}
\end{figure*}

\begin{figure}[thb]
    \centering
    \includegraphics[width=0.45\textwidth]{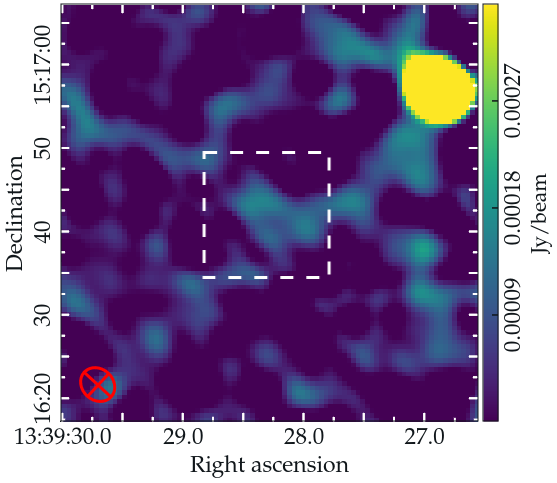}
    \caption{The non-detection of the source GP3. The white dashed square is centred on the optical coordinates of the source with \(15\times15''\) size. The red ellipse indicates the beam size and shape. The source is not detected due to its faintness and higher RMS noise in the image near the source coordinates.}
    \label{fig:enter-label}
\end{figure}

We selected the three GPs with X-ray observations from \citet{Svoboda19} for targeted JVLA observations (ID: VLA/21B-136, PI: A. Borkar). All observations were taken in the B-configuration. Continuum observations were scheduled in L-band ({1 $-$ 2 GHz with the effective frequency (EF) of} 1.4 GHz), C-band ({4 $-$ 8 GHz, EF} 6 GHz) and X-band ({8 $-$ 12 GHz, EF} 10 GHz) to be able to construct a radio spectral energy distribution (SED). Eight out of the scheduled nine blocks were observed, with \(\sim 14\) minute on source time and \(\sim 30\) minutes total time including calibration observations. 3C147 was used as a bandpass and amplitude calibrator, and J0741+3112 as the phase calibrator. The observations were performed with a short bandpass calibration scan at the beginning. The targets were observed for two sets of \(\sim 7\) minutes on-source, bracketed by phase calibration scans. GP1 L-band observation was not completed during the observing period due to low priority. The GP3 observations were performed with incorrect target coordinates, and thus, the source is outside the field of view for the C- and X-band observations, while it is still within the field of view in the L-band, but close to the edge of the primary beam, where the RMS noise is higher.

The data reduction and imaging was performed using the standard calibration pipeline with the Common Astronomy Software Applications (CASA v6.2.1, \citealt{CASA22}) package. We used the multi-frequency synthesis and Briggs weighting with robust parameter = 0.5 for the imaging. The synthesised beam (angular resolution) in the B-configuration at L-, C- and X-bands is 4.3, 1.0 and 0.6 arcsec. The targets remain unresolved at this resolution. A summary of the observations {with confirmed detections} is provided in Table \ref{tab:VLA_obs}.

The radio images of the three GPs are shown in Fig.~\ref{fig:GP-obs}. Sources GP1 and GP2 are detected in both C and X bands and have comparable fluxes in both the bands. The RMS noise in these images is \(\sim 7~\mu\)Jy. The L-band observation of GP1 was not performed during the observation cycle, and the GP2 observation in the L-band is contaminated by the artefacts from a particularly bright off-axis radio source. The radial streaks from the contaminating source pass close to the position of GP2, rendering it undetectable, and could not be mitigated by self-calibration.
The source GP3 is within the field of view in the L-band, but there is no significant detection. The RMS noise near the source position is \(\sim 70~\mu\)Jy due to its location close to the edge of the field of view, thus the signal-to-noise is not sufficient to detect a \(\leq 100~\mu\)Jy source. In Table~\ref{tab:VLA_obs}, we present the observation and source parameters for the detected sources.

\subsection{Radio data from archival surveys}

We supplement our targeted JVLA observations with the images from all-sky radio surveys with multiple instruments at different frequencies. We used the data from the TIFR GMRT Sky Survey (TGSS, \citealt{TGSS}), an all-sky survey at 150 MHz with the Giant Metrewave Radio Telescope (GMRT), the VLA surveys (1) the Faint Images of the Radio Sky at Twenty-centimeters (FIRST) Survey \citep{FIRST} and (2) the NRAO VLA Sky Survey (NVSS, \citealt{NVSS}) at 1.4 GHz, (3) the VLA Sky Survey (VLASS, \citealt{VLASS}) at 3 GHz; the LOFAR Two-metre Sky Survey (LoTSS, \citealt{LoTSS-DR2}) with the Low Frequency Array at 150 MHz, and the Rapid ASKAP Continuum Survey first data release at 888 MHz (RACS, \citealt{RACS-low}), which provides a coverage for the sources in the southern sky. The properties of these surveys are provided in the Table~\ref{tab:archival_surveys}.

\begin{table}
    \begin{center}
        \caption{Summary of the archival radio surveys used for flux measurements of the targeted and archival sources.}
        \label{tab:archival_surveys}
        \begin{tabular}{lclccc}
            \toprule
            \toprule
            Survey & Observatory & Freq. & resolution & sensitivity\\
            & & [GHz] & [arcsec] & [mJy\,beam\(^{-1}\)]\\[1mm]
            \midrule\\[-3mm]
            FIRST & VLA & 1.4 & 5.4 & 0.15\(^{a}\)\\
            NVSS & VLA & 1.4 & 45 & 0.45\(^{a}\) \\
            VLASS & VLA & 3.0 & 2.5 & 0.145\\
            TGSS & GMRT & 0.15 & 25 & 5 \\
            LoTSS & LOFAR & 0.15 & 6 & 0.083\\
            RACS & ASKAP & 0.888 & 15 \textendash~25 & 0.2\\[1mm]
            \bottomrule
        \end{tabular}
    \end{center}
Note: \((a)\) Although the median RMS noise is lower, the source detection limit is \(5 \times rms\) and the catalogues contain sources above this limit.

\end{table}

\begin{figure}[ht]
    \centering
    \includegraphics{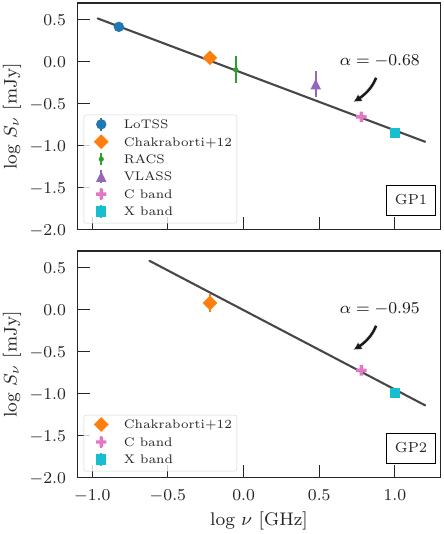}
    \caption{The radio SED of GP1 and GP2. The data points are from the JVLA observations at 6 and 10 GHz, and archival measurements. GP2 shows a steeper spectral index compared to GP1. GP3 was not detected in any archival surveys.}
    \label{fig:GP-SEDs}
\end{figure}

\begin{table*}
\begin{center}
    \caption{Archival detections of the targeted GPs with expected and observed flux densities in mJy. The expected fluxes are derived from Eqs. \ref{eq:1}, \ref{eq:2} \& \ref{eq:3}.}
    \label{tab:archival-targeted-GPs}
    \begin{tabular}{lcccccccccc}
        \toprule
        \toprule
        GP & \multicolumn{2}{c}{LoTSS\(^{(a)}\)} & \multicolumn{2}{c}{GMRT\(^{(b)}\)} & \multicolumn{2}{c}{RACS\(^{(c)}\)} & \multicolumn{2}{c}{NVSS/FIRST\(^{(d)}\)} & \multicolumn{2}{c}{VLASS\(^{(e)}\)} \\
        name & \multicolumn{2}{c}{[0.15 GHz]} & \multicolumn{2}{c}{[0.62 GHz]} & \multicolumn{2}{c}{[0.88 GHz]} & \multicolumn{2}{c}{[1.4 GHz]} & \multicolumn{2}{c}{[3 GHz]} \\
         & Exp. & Obs. & Exp. & Obs. & Exp. & Obs.& Exp. & Obs. & Exp. & Obs.\\
        \midrule
        GP1 & $2.3 - 3.9$ & 2.6 & $0.8 - 1.3$ & 1.1 & $0.7 - 1.0$ & 0.8 & $0.5 - 0.7$& <0.45 & $0.3 - 0.5$ & 0.54 \\
        GP2 & $2.4 - 4.1$ & --- & $0.9 - 1.5 $ & 1.2 & $0.7 - 1.1$ & <0.2 & $0.5 - 0.8$ & <0.45 & $0.3 - 0.5$ & <0.15 \\
        GP3 & $1.6 - 2.6$ & --- & $0.6 - 0.9$ & --- & $0.5 - 0.7$ & <0.2 & $0.3 - 0.5$ & <0.45 & $0.2 - 0.3$ & <0.15 \\
        \bottomrule
    \end{tabular}
\end{center}

\footnotesize{\((a)\) LoTSS: \citet{LoTSS-DR2},
\((b)\) GMRT: \citet{Chakraborti12},
\((c)\) RACS: \citet{RACS-low},
\((d)\) NVSS: \citet{NVSS}, FIRST: \citet{FIRST},
\((e)\) VLASS: \citet{VLASS}}.
\end{table*}

We searched the all-sky surveys mentioned in Table~\ref{tab:archival_surveys} to look for the detection of our GP sample at different frequencies in order to build their radio SEDs. In the archival data, GP1 was strongly detected in the LoTSS catalogue with a flux of 2.6 mJy, although it may also be affected by artefacts from a bright off-axis source. In the RACS and the VLASS catalogues, GP1 was detected at 0.8 mJy and 0.54 mJy, respectively, but the detections are below the \(5\sigma\) limit of those surveys. GP1 is unresolved in all archival data, as the surveys lack the necessary resolution. GP2 and GP3 were undetected in all the survey catalogues. LoTSS, which has the best sensitivity among the aforementioned catalogues, currently does not have a sky coverage below a declination of +28\(^\circ\) and thus has not observed GP2 and GP3 yet, but may in the future data releases be able to detect these sources.

A sub-sample of the \citet{Cardamone09} GPs was observed by \citet{Chakraborti12} with GMRT at 617 MHz. They observed three GP sources \textemdash~including GP1 and GP2 \textemdash~which were expected to have a \(\sim\)1 mJy flux at 617 MHz. The third source, J142405.73+421646.3, though different from our GP3, has similar galaxy stellar mass (\(\Mstar\sim 9\)) and SFR (\(19.6~\mathrm{M_\odot\,yr^{-1}}\)) and is located at a comparable redshift of 0.18 as GP3. Both GP1 and GP2 were detected at 1 mJy level, while the third source was undetected, similar to GP3. In Table~\ref{tab:archival-targeted-GPs}, we provide the detected flux densities of our targeted GPs in archival observations. The expected flux density ranges are estimated from the theoretical and empirical relations between SFR and radio luminosities (Eqs.~\ref{eq:1},~\ref{eq:2} and \ref{eq:3}, see Sec.~\ref{sec:SFR-relation} for details). The detections are consistent with the estimates from the SFR relations.

The detected fluxes of the two GPs were used to construct their radio SEDs, which are shown in Fig.~\ref{fig:GP-SEDs}. GP1 shows a standard power-law SED with the spectral index of \(\alpha = -0.68 \pm 0.01\). GP2 shows an SED with a steeper power law with the spectral index of \(\alpha=-0.95 \pm 0.17\). The spectral indices are characteristic of synchrotron emission, which can arise from star forming activity as well as from accretion onto black holes \citep{Panessa19}.

\section{Comparison with the dwarf galaxy population}\label{sec:Comparison}

The detection of radio emission in two GP galaxies and the non-detection in one (or two, if we also include the third source from \citealt{Chakraborti12}) raises the question of whether these detections are exceptional or representative of GP-like star forming galaxies. To test this, we searched for the radio counterparts for the GP and BB samples in the archival radio catalogues mentioned in Sect.~\ref{sec:Data Analysis}. We cross-matched the GP sample from \citet{Cardamone09} (80 star-forming and 8 narrow-line Seyfert 1 (NLS1) sources), along with the GPs from \citet{Keel22} (38 sources) and \citet{Brunker22} (13 sources), and the BB sample from \citet{Yang17} (40 sources), \citet{Jaskot19} (13 sources) and \citet{McKinney19} (17 sources). The radio catalogue images were further visually inspected to verify the detection (or the lack thereof).

\begin{table*}[ht]
\begin{center}

    \caption{Radio detected archival GP and BB galaxies.}
    \label{tab:archival-GPs-BBs}
    \begin{tabular}{lccccccccc}
        \toprule
        \toprule
        Source & RA & DEC & \(z\) & LoTSS & TGSS & NVSS & FIRST & VLASS & Classification\\
         id & [h:m:s] & [d:m:s] & & [mJy] & [mJy] & [mJy] & [mJy] & [mJy] & \\[1mm]
         \midrule\\[-3mm]
        C09 15 & 14:11:45.327 & +62:39:11.212 & 0.2301 & 0.5 & --- & --- & --- & --- & SF\\
        C09 17 & 15:40:50.206 & +57:24:41.935 & 0.2944 & 0.7 & --- & --- & --- & --- & SF\\
        C09 20 & 07:49:36.773 & +33:37:16.395 & 0.2733 & 2.6 & --- & --- & --- & 0.54 & SF\(^{\dag}\)\\
        C09 25 & 09:26:00.403 & +44:27:36.171 & 0.1807 & 0.7 & --- & --- & --- & --- & SF\\
        C09 26 & 13:01:28.312 & +51:04:51.200 & 0.3479 & 6.1 & --- & --- & --- & 0.7 & QC\(^{\ddag}\)\\
        C09 31 & 10:53:30.815 & +52:37:52.858 & 0.2526 & 1.2 & --- & --- & --- & --- & SF\\
        C09 32 & 08:47:54.082 & +33:36:54.646 & 0.3063 & 1.3 & --- & --- & --- & --- & SF\\
        C09 33 & 13:39:40.709 & +55:27:40.100 & 0.2291 & 0.7 & --- & --- & --- & --- & SF\\
        C09 44 & 09:57:39.774 & +37:42:07.581 & 0.2867 & 0.5 & --- & --- & --- & --- & SF\\
        C09 45 & 14:40:09.956 & +46:19:36.970 & 0.3008 & 1.6 & --- & --- & --- & --- & SF\\
        C09 46 & 14:54:35.581 & +45:28:56.248 & 0.2687 & 0.8 & --- & --- & --- & --- & SF\\
        C09 58 & 11:37:22.138 & +35:24:26.623 & 0.1945 & 0.5 & --- & --- & --- & --- & SF\\
        C09 59 & 11:26:37.771 & +38:03:03.062 & 0.2469 & 0.5 & --- & --- & --- & --- & SF\\
         & & & & & & & & & \\
        C09 N2 & 14:19:18.900 & +51:02:40.142 & 0.3236 & 6.8  & --- & --- & 1.45 & 1.4 & NLS1 \\
        C09 N3 & 16:22:09.407 & +35:21:07.269 & 0.2660 & 13.2 & --- & --- & 2.0 & 1.8 & NLS1 \\
        C09 N4 & 07:49:32.947 & +28:34:06.750 & 0.3369 & 342  & 390 & 27 & 28 & 10.2 & NLS1 \\
        C09 N5 & 11:29:07.104 & +57:56:05.206 & 0.3123 & 11.4 & --- & --- & 1.92 & 1.8 & NLS1 \\
        C09 N6 & 11:26:15.265 & +38:58:17.389 & 0.3365 & 2.2  & --- & --- & 1.2 & --- & NLS1 \\
        C09 N7 & 08:18:00.198 & +19:18:10.181 & 0.3245 & ---\(^*\)  & --- & --- & 1.4 & 0.93 & NLS1\\
         & & & & & & & & & \\
        K22 1 & 15:04:57.987 & +59:54:07.27 & 0.2502 & 0.6 & --- & --- & --- & --- & SF\\
        K22 4 & 14:10:05.248 & +53:50:37.89 & 0.3353 & 0.5 & --- & --- & --- & --- & SF\\
        K22 20 & 13:01:28.316 & +51:04:51.18 & 0.3479 & 6.1 & --- & --- & --- & 0.7 & QC\(^{\ddag}\)\\
         & & & & & & & & & \\
        B22 1 & 13:10:09.40 & +29:17:42.9 & 0.3582 & 0.5 & --- & --- & --- & --- & SF\\
        B22 5 & 12:22:24.2 & +43:11:23.0 & 0.3049 & 0.4 & --- & --- & --- & --- & SF\\
        B22 6 & 13:14:41.28 & +43:43:26.5 & 0.2939 & 0.68 & --- & --- & --- & --- & SF\\
        B22 7 & 13:15:49.44 & +43:34:30.4 & 0.3276 & 0.7 & --- & --- & --- & --- & SF\\
        B22 9 & 14:29:40.9 & +43:54:09.0 & 0.3592 & 0.7 & --- & --- & --- & --- & SF\\
        B22 10 & 14:36:20.3 & +43:53:02.0 & 0.3311 & 1.3 & --- & --- & --- & --- & SF\\
        B22 11 & 15:26:23.8 & +43:00:16.45 & 0.3688 & 1.5 & --- & --- & --- & --- & SF\\
        B22 12 & 15:42:46.5 & +43:53:58.20 & 0.3080 & 0.5 & --- & --- & --- & --- & SF\\
         & & & & & & & & & \\
        Y17 8 & 15:09:34.174 & +37:31:46.12 & 0.0326 & 0.4 & --- & --- & --- & --- & SF\(^{\S}\)\\
         & & & & & & & & & \\
        J19 4 & 08:51:15.655 & +58:40:54.997 & 0.0919 & 0.5 & --- & --- & --- & --- & SF\\
        J19 10 & 15:09:34.174 & +37:31:46.12 & 0.0325 & 0.4 & --- & --- & --- & --- & SF\(^{\S}\)\\
        J19 12 & 17:35:01.229 & +57:03:08.458 & 0.0472 & ---\(^*\) & --- & --- & 1.3 & 1.2 & AGN\\
         \bottomrule
    \end{tabular}
\end{center}
Notes: \(^*\):= Not observed; C09:=\citet{Cardamone09}, K22:=\citet{Keel22}, B22:=\citet{Brunker22}, Y17:=\citet{Yang17}, J19:=\citet{Jaskot19}. SF = star forming; QC = quasar candidate \citep{SDSS-DR7}; NLS1 = narrow line Seyfert 1; AGN = AGN candidate \citep{Sartori15}. \dag: GP1; \ddag: C09 26 same as K22 20; \S: Y17 8 same as J19 10.

\end{table*}

The list of GP and BB sources with a radio counterpart detection is provided in the Table~\ref{tab:archival-GPs-BBs}. Older all-sky surveys, viz. FIRST, NVSS and TGSS, have significantly lower sensitivity and thus the GP and BB sources would not be detected in these surveys. None of the star-forming GP or BB sources were detected at \(>5\sigma\) level in the VLASS or RACS surveys, which have slightly better sensitivity. For the LoTSS survey, the majority of the sources lie outside the current footprint and are not observed. Still, thanks to its excellent sensitivity and lower frequency where the sources are likely to be brighter, several sources are detected. In total, 13 star-forming GPs from \citet{Cardamone09}, 8 from \citet{Brunker22}, 3 from \citet{Keel22}, 1 BB from \citet{Yang17} and 3 BBs from \citet{Jaskot19} are detected.

However, 42 GPs and 17 BBs were observed within the LoTSS footprint without any detection, giving a detection rate of 35 and 17 percent respectively. A sub-sample of 10 \citet{Yang17} BBs was observed by \citet{Sebastian19} using GMRT at 1.4 GHz, where they found 9 out of 10 detections. Most of the sources in their sample are considerably weak with flux densities \(\leq 0.1\) mJy. Only two sources have flux \(>0.1\) mJy, one of which is within the LoTSS footprint, which is the only BB from \citet{Yang17} detected there. A simple two-point SED gives a very flat spectral index of \(\alpha = -0.044\), which is often a characteristic of thermal free-free emission or free-free absorption, which trace the current star formation \citep{Clemens10}.

Along with the star-forming GPs, \citet{Cardamone09} also identified 8 NLS1 galaxies, out of which, 5 were detected in LoTSS (and the other 3 are not observed), 4 in VLASS, and 1 in TGSS \& NVSS. All the sources are detected with strong radio emission ranging from 2 to 400 mJy at low frequency (LoTSS and TGSS) and \(\geq 1\) mJy at high frequency in VLASS. Their radio spectral indices \(\alpha\) lie between \(-0.56\) and \(-0.8\), which are typical for radio-loud AGN. This is in contrast with the radio detections of star-forming GPs, where the total flux is \(\leq 1\) mJy at low frequency in LoTSS and undetected in other surveys. The two exceptions to this are (1) 195.36801+51.080893 (C09 26 in Table~\ref{tab:archival-GPs-BBs}) which is detected at 6.1 mJy and is considered as a quasar candidate \citep{SDSS-DR7}. The LoTSS image of this source shows an elongated radio structure which could be due to jet emission. This source is also close to (\(<25''\)) another strong radio source, the radio emission associated with C09 26 may be contaminated by or associated with a radio jet from this nearby source. (2) 117.403215+33.621219, i.e. GP1 (C09 20 in Table~\ref{tab:archival-GPs-BBs}), which is detected at 2.6 mJy. Similarly, one BB from the \citet{Jaskot19} sample (J19 12 in Table~\ref{tab:archival-GPs-BBs}) is detected in VLASS and FIRST at \(\sim 1.2\) mJy. This source is identified as an AGN candidate using a mid-IR colour-cut (see \citealt{Sartori15,Comerford20} for details). The source shows a change in flux between different VLASS epochs (1.6 mJy in epoch 1.1, 1.2 mJy in epoch 2.1 and 1.0 mJy in epoch 3.1), suggesting that it may be variable.

In summary, a typical star-forming GP or BB galaxy is likely radio-weak or undetected with its low frequency flux \(\leq 1\) mJy, while the galaxies hosting AGN or quasar candidates show strong radio emission and are detected in multiple surveys. With a strong radio detection at several radio frequencies, GP1 shows a closer similarity to the Seyfert galaxies than to the star-forming GPs, suggesting that it may host an AGN.

\section{Star formation rate and radio flux relation}\label{sec:SFR-relation}

We test whether the radio emission from the targeted GPs and archival sources is consistent with the expected emission from star formation, and investigate how many sources would be detectable with the archival surveys. The radio emission in star-forming galaxies arises from synchrotron emission from the relativistic electrons accelerated from supernova explosions interacting with the galactic magnetic field, and thermal bremsstrahlung (free-free) emission from ionised hydrogen gas surrounding hot young stars. At lower radio frequencies (\(\leq 10\) GHz), the synchrotron emission is dominant over the free-free emission \citep{Condon92}, and we can estimate the expected radio emission arising from SFR using only the synchrotron emission. \citet{Yun02} provide a theoretical expected non-thermal synchrotron flux density for a galaxy as
\begin{equation}
    S(\nu) = 25f_{nth}\nu^{\alpha}\,\mathrm{SFR}\,D_L^{-2}~[\mathrm{Jy}]\label{eq:1}
\end{equation}
where \(\alpha\) is the synchrotron spectral index (\(\sim -0.7\)) and \(D_L\) is the luminosity distance in Mpc, \(f_{nth}\) is a scaling factor introduced to account for the variations in normalisation. The Table 1 in \citet{Yun02} provides the value of \(f_{nth}\) for some nearby galaxies, which are of the order \(\sim1\). We take the mean of this distribution (\(f_{nth} = 1.35\)) as the value for our calculations, and use the standard deviation (\(\sigma_{f} = 0.95\)) as the error-range.

For an empirical relation, \citet{Bell03} provide an estimation of SFR using far-IR and UV luminosities to calibrate the radio emission:
\begin{equation}
    \mathrm{SFR} = 5.52 \times 10^{-22}~L_{1.4\,\mathrm{GHz}}\label{eq:2}
\end{equation}
for bright starburst galaxies. We invert this relation, using the SFR obtained from UV-optical observations, to estimate the radio luminosity and in turn, the radio flux density at 1.4 GHz, and extrapolate to different frequencies assuming \(\alpha = -0.7\). Both Eq.~\ref{eq:1} and Eq.~\ref{eq:2} were used for estimating the range of expected radio flux densities and the detectability of the sources at higher frequencies.

The largest number of sources are detected in the LoTSS survey. Most of the radio-to-SFR relations are calibrated to 1.4 GHz frequency or higher, and may not accurately reflect the expected flux at low frequency at 150 MHz. Thus, we use the relation between low-frequency radio flux and SFR derived for the LOFAR data, usually expressed as:
\begin{equation}
    \mathrm{log}(L_{150}) = a_1 \,\mathrm{log(SFR)} + a_2 \label{eq:3}
\end{equation}
where \(L_{150}\) is the 150 MHz luminosity (in units of W\,Hz$^{-1}$) detected with LOFAR. \citet{Gurkan18} used the SDSS-selected sample of 2000+ galaxies to study \(L_{150}-\mathrm{SFR}\) relation, and find the best-fit values of \(a_1=1.07\pm0.01\) and \(a_2=22.06\pm0.01\), while \citet{Smith21} used more than 100,000 near-IR selected sources up to z\(\sim 1\) and find similar values of \(a_1=1.041\pm0.007\) and \(a_2=22.181\pm0.005\). More recently, using 45 nearby galaxies (\(<30\) Mpc), \citet{Heesen22} found a steeper relation with \(a_1=1.4\pm0.07\) and \(a_2=21.57\pm0.04\). Using these relations, we calculate the expected luminosities for the GP and BB sources at 150 MHz.

\begin{figure}[htb]
    \centering
    \includegraphics[width=0.46\textwidth]{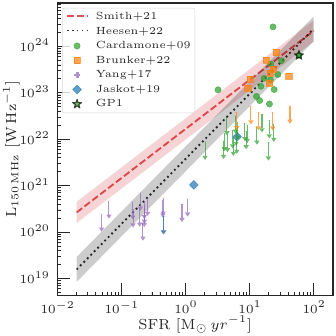}
    \caption{The 150 MHz luminosities of the star-forming galaxies observed in the LoTSS survey as a function of their SFR. The green circles and orange squares show the detected GPs while the purple crosses and blue diamonds show the detected BBs. The sources not detected in the LoTSS survey are shown by arrows as upper limits. The red dashed and the black dotted lines show the expected luminosity for a given SFR using the \citet{Smith21} and \citet{Heesen22} relations, resp., and the shaded regions represent the scatter in the relations. The luminosities of most detected sources are either consistent or underluminous with respect to SFR. All the undetected sources are significantly underluminous.}
    \label{fig:Rlum-SFR}
\end{figure}

The detected flux densities from GP1 and GP2 are overall consistent with the expectations from the SFR relation. As discussed in Sec.~\ref{sec:Comparison}, most of the GPs \& BBs appear to be radio-weak (\(\leq 1\) mJy for the detected and \(<0.1\) mJy for the undetected sources), thus it is useful to put the fluxes of our targeted GPs in this context and estimate their significance. Comparing the expected fluxes from the given SFR for the star-forming GPs and BBs at the different frequencies for respective catalogues, we find that none of the sources would be detectable at \(\geq 5\sigma\) level with the TGSS, FIRST, NVSS, or VLASS surveys. Two sources in this sample, viz. GP1 and GP2, have the expected 3 GHz flux from Eqs. (1) and (2), which would be detectable with VLASS at \(3\sigma\) and \(2\sigma\) level, respectively. Following the \citet{Smith21} relation (Eq.~\ref{eq:3}), we expect \(\sim 75\%\) (70/93) star-forming GPs and \(\sim 83\%\) of BBs (44/53) to be detectable at \(>5\sigma\) level, but only 47\% (21/45) GPs and 10\% (2/19) BBs are actually detected. Similarly, the \citet{Heesen22} relation predicts \(\sim 57\%\) (53/93) GPs and \(\sim 40\%\) (21/53) BBs could be detected. In Fig.~\ref{fig:Rlum-SFR}, we plot the observed 150 MHz luminosities of the star-forming GPs and BBs as a function of their SFR, together with the expected luminosity following the \citet{Smith21} relation (red dashed line), and the \citet{Heesen22} relation (black dotted line). The GPs which have been detected lie on the relations within the error margin, or are below the relation by a factor of a few. One exception to this is the source which is a quasar candidate (C09 26/K22 20 in Table~\ref{tab:archival-GPs-BBs}), and is overluminous by a factor of \(\sim 7\). The source GP1 has 150 MHz luminosity about the same as expected from the SFR. The high radio flux detected for GP1 compared to the rest of the GP sample is likely due to its high SFR, which is the highest SFR among all GPs and BBs considered in this work. The two detected BBs are below the \citet{Smith21} relation by an order of magnitude. All the undetected sources are also well below this relation. The steeper \(L_{150}-\)SFR relation by \citet{Heesen22} reduces the discrepancy between the observed and the expected luminosities. But the detected two BBs and the undetected sources are below the relation by a factor of few. The \citet{Gurkan18} and \citet{Smith21} relations are based on statistically significant large samples up to \(z\leq1\) spanning a wide range of galaxy SFRs that are representative of the GP and BB galaxies. The galaxies in \citet{Heesen22} sample are much closer (\(z<0.01\)) and thus weaker sources are easier to detect, which could steepen the relation. All three relations are based on galaxies with larger stellar masses (\(\Mstar \geq 8.5\)), which cover the GP mass range, but are more massive than the BBs (\(\Mstar \approx 6-8\)). \citet{Gurkan18} and \citet{Smith21} suggest that the \(L_{150}-\)SFR also depends on the galaxy stellar mass: \(\mathrm{log}(L_{150}) = a_1\,\mathrm{log(SFR)} + a_2\,\mathrm{log(M_\star)} + a_3\). Using the values for the coefficients \(a_1\), \(a_2\) and \(a_3\) provided by these authors, we compare the estimated and observed luminosity. The observed luminosities of the detected GPs are consistent with or higher than expected luminosities, while the detected BB and the undetected sources have observed luminosities below the expectation by a factor of few. Thus, even accounting for the mass-dependence, lower-mass galaxies appear underluminous. Our analysis suggests that the \(L_{150}-\)SFR relation is possibly even steeper towards the lower-mass end, and a careful analysis using a sample consisting of low-mass galaxies is needed to establish accurate relations spanning the whole galaxy mass range.

\section{Conclusions and Discussion}\label{sec:Conclusions}
In this paper, we present the JVLA observations of three GP galaxies.

\begin{enumerate}

\item \textbf{JVLA Observations:}
GP1 and GP2 are clearly detected in C- and X-band observations. GP1 L-band observation was not performed and GP2 L-band image is affected by artefacts from a nearby bright source. GP3 was undetected in the L-band observations, while it is outside the field of view in the C- and X-band images.\vspace{1mm}

\item \textbf{Archival observations:} We supplemented our JVLA observations with archival radio surveys to construct radio SEDs for GP1 and GP2. The detected radio fluxes of the two sources are consistent with the expected radio emission from star formation. GP3 is not detected in any archival surveys.\vspace{1mm}

\item \textbf{Supplementary sample:} We compare our observations with a larger sample of GPs and BBs in the archival surveys and in literature to understand the nature of radio emission in dwarf star-forming galaxies. We find that only a small fraction (36\%) of GPs and BBs are detected in the archives. On the other hand, dwarf galaxies which have been classified as AGNs or AGN candidates show strong radio emission and are often detected in several archival surveys.\vspace{1mm}

\item \textbf{Radio flux-SFR relation:} We compare the detected radio luminosity of the star-forming galaxies with the luminosity expected from SFR derived from empirical relations. Although about \(55-75\)\% of sources could be detected with the most sensitive LoTSS survey, fewer than 40\% are detected. A few of the detected and most of the undetected sources are underluminous compared to the expectations from the empirical relations. This suggests that towards the lower end of the galaxy mass and SFR, the radio flux-to-SFR relation deviates away from the relations established for larger galaxies. Typical compact, highly star-forming dwarf galaxies, such as GPs and BBs are markedly underluminous, and the presence of strong radio emission may hint at an alternative source such as an AGN.

\end{enumerate}

Such a deficiency in the radio flux of GPs and BBs has been reported before. \citet{Chakraborti12} have shown that GPs have lower radio flux in comparison with local starburst galaxies and expected from empirical relations. \citet{Sebastian19} show that the radio derived SFR is significantly less than the H\(\alpha\)-derived SFR. The radio flux in the star-forming galaxies likely depends on the galaxy mass, as \citet{Kouroumpatzakis21} have shown that there is a significant deficit in radio luminosity compared to H\(\alpha\) luminosity for low mass galaxies. GPs and BBs, with \(\Mstar\sim 9\) and \(\Mstar\sim 7\), respectively, are on the lower end of the mass range and show a similar trend. Our analysis with a large sample of low-mass, high-SFR galaxies confirms that they do not follow the same relation as high-mass galaxies.

The radio and IR emission in star-forming galaxies is often described by the ``calorimeter'' model \citep{Voelk89}. The energy from UV photons is completely reprocessed by the dust in the galaxy and radiated in the far-IR, while the energy from electrons driven by supernova explosions is extracted entirely as synchrotron emission before they escape the galaxy. Thus, the galaxies behave as both UV and electron calorimeters. Several studies have now shown that towards the low mass end, galaxies may not act as perfect calorimeters and both the radio and IR emission is suppressed \citep{Bell03,Boyle07,Beswick08,Roychowdhury12,Smith21,Heesen22,Kouroumpatzakis23}, as the dust content and the efficiency of synchrotron emission is reduced by a similar factor \citep{Lacki10a,Lacki10b}. Our analysis demonstrates that towards the lowest end of galaxy mass and SFR, the radio emission deviates from the radio-SFR relation established for more massive galaxies, supporting the previous studies. This could be due to a higher escape fraction of synchrotron emitting electrons due to a weaker gravitational potential of the low-mass galaxies. Dwarf galaxies also do not show settled morphology, and consequently they may not host large-scale and high-intensity magnetic fields.

The radio observations of the three targeted GPs show similarities with their X-ray observations, where GP1 and GP2 were detected in the X-rays with more than 5 times excess flux compared to what is expected from SFR relations, while GP3 remained undetected \citep{Svoboda19}. This X-ray excess cannot be explained by a higher number of high mass X-ray binaries or excess emission from the hot gas \citep{Franeck22} and is often attributed to the presence of an AGN. The X-ray and radio emission could arise from the same source, which is different from the emission in the optical band. A systematic comparison of the multi-wavelength properties of star-forming dwarf galaxies is necessary to understand the cause of the emission (or the lack of detectable emission) and the relation between the radio and X-ray regimes, and how it differs from the optical and IR emission. Surveys from recently launched and next generation of observatories (e.g \textit{JWST, Euclid, Athena}, the Square Kilometre Array, Vera C. Rubin Observatory etc.) will be able to access the areas of parameter space occupied by lower-mass and dwarf galaxies, even at higher redshift. Thus, it is crucial to understand how the luminosity-SFR relations hold for the low-mass galaxies with a diverse range of SFRs and cosmic ages.

\begin{acknowledgements}
We thank the anonymous referee for their helpful comments which helped to improve the manuscript.

This work was supported from the Czech Science Foundation project No.22-22643S. BM acknowledges support from the Science and Technology Facilities Council (STFC) under grants ST/T000295/1 and ST/X001164/1.

This publication makes use of data products from NSF's Karl G. Jansky Very Large Array (VLA). The National Radio Astronomy Observatory is a facility of the National Science Foundation operated under cooperative agreement by Associated Universities, Inc.

This research has made use of the CIRADA cutout service at URL cutouts.cirada.ca, operated by the Canadian Initiative for Radio Astronomy Data Analysis (CIRADA). CIRADA is funded by a grant from the Canada Foundation for Innovation 2017 Innovation Fund (Project 35999), as well as by the Provinces of Ontario, British Columbia, Alberta, Manitoba and Quebec, in collaboration with the National Research Council of Canada, the US National Radio Astronomy Observatory and Australia’s Commonwealth Scientific and Industrial Research Organisation.

This research has made use of ESASky \citep{ESASky1,ESASky2}, developed by the ESAC Science Data Centre (ESDC) team and maintained alongside other ESA science mission's archives at ESA's European Space Astronomy Centre (ESAC, Madrid, Spain).

This research has made use of the SIMBAD database, operated at CDS, Strasbourg, France \citep{Simbad}.

Software used: \texttt{Numpy} \citep{Numpy}, \texttt{matplotlib} \citep{Matplotlib}, \texttt{astropy} \citep{astropy:2013,astropy:2018,astropy:2022}, \texttt{CARTA} \citep{CARTA}.
\end{acknowledgements}

%
%
\bibliographystyle{aa}
\bibliography{References}

\end{document}